\documentclass[5p,12pt,a4paper,sort&compress]{elsarticle}
\usepackage{natbib}
\usepackage{textcomp}
\usepackage{dcolumn}
\usepackage{bm}
\usepackage{datetime}
\usepackage[colorinlistoftodos]{todonotes}         % display notes
\usepackage[pdftex]{hyperref}
\usepackage{soul}
\usepackage{nomencl}
\usepackage{mathrsfs}
\usepackage{setspace}
\usepackage{svn-multi}
\usepackage{amsmath}
\usepackage{amsfonts}
\usepackage{amssymb}
\usepackage{lineno}
\usepackage[nomarkers]{endfloat}
\newcounter{todoListItems}

\def\mys#1{{\mbox{\scriptsize{#1}}}}    % font for capital subscripts

\newcommand{\degc}{\ensuremath{^{\circ}}\mathrm{C}}
\def\rg{r_{\mys{g}}}            % polymer radius of gyration
\def\a{a}                        % Particle radius
\def\cp{c_{\rm{p}}}             % polymer concentration
\def\cpstar{c_{\rm{p}} / c_{\rm{p}}^{*} }             % scaled polymer concentration
\def\phic{\phi_{0}}             % Initial colloid volume fraction
\def\phicolloid{\phi}           % colloid volume fraction
\def\Uc{\U_{\mys{c}}}            % potential at contact
\def\U{U}                       % interparticle potential
\def\D{\Delta}                  % range of potential
\def\etaL{\eta_{\mys{L}}}       % low-shear limiting viscosity
\def\R{\xi}            % Mean cluster size 
\def\h0{h_{0}}                  % initial height of gel
\def\hcrit{h_{\mys{crit}}}      % critical height of gel
\def\lcrit{l_{\mys{crit}}}      % stress transmission length
\def\lg{\left < l_{\mys{s}} \right >  }               % mean width of strands
\def\taud{\tau_{\mys{d}}}           % delay time
\def\taumean{\left < \taud \right >} % mean delay time
\def\taulife{\tau_{\mys{N}}}    % lifetime of percolating network of gel
\def\taubreak{t_{\mys{break}}} % stress-dependent time before gel yields
\def\tw{t_{\mys{w}}}             % waiting time
\def\tauesc{\tau_{\mys{esc}}}   % bond lifetime
\def\tauzero{\tau_{0}}             % characteristic time
\def\vs{\nu_{\mys{s}}}           % slow initial settling rate (poroelastic) of gels
\def\vf{\nu_{\mys{f}}}           % fast initial settling rate  of gels in collapse region
\def\v0{\nu_{0}}                 % settling velocity of particle at infinite dilution
\def\sigg{\sigma_{\mys{B}}}     % Buoyant stress on gel
\def\sigcrit{\sigma_{\mys{crit}}} % critical stress
\def\sigyield{\sigma_{\mys{Y}}} % compressive yield stress
\def\k0{k_{0}}                  % initial permeability of gel
\def\kb{k_{\mys{B}}}      %\noindent Revision: \svnrev \\
\def\kBT{\kb T}                 % Thermal energy
\def\NA{N_{\mys{A}}}            % Avogadro's constant
\begin{document}

\svnidlong
{$HeadURL: https://macaulay.dyndns-server.com/svn/Delay-paper-2014/delay.tex $}
{$LastChangedDate: 2014-02-20 18:25:55 +0000 (Thu, 20 Feb 2014) $}
{$LastChangedRevision: 80 $}
{$LastChangedBy: admin $}
\svnid{$Id: delay.tex 80 2014-02-20 18:25:55Z admin $}

\bibliographystyle{elsarticle-num-names}
\title{Gels under stress: the origins of delayed collapse}

\author[uob]{Lisa J.~Teece}
\address[uob]{School of Chemistry, University of Bristol, Bristol
BS8 1TS, United Kingdom}
\author[uob]{James M.~Hart}
\author[uob]{Kerry Yen Ni Hsu}
\author[uob]{Stephen Gilligan}
\author[BCS]{Malcolm A.~Faers}
\address[BCS]{Bayer CropScience AG, 40789, Monheim am Rhein, Germany}
\author[uob]{Paul~Bartlett\corref{cor1}}
\ead{p.bartlett@bristol.ac.uk}
\cortext[cor1]{Corresponding author}

\begin{abstract}

Attractive colloidal particles can form a disordered elastic solid or gel when quenched into a two-phase region, if the volume fraction is sufficiently large. When the interactions are comparable to thermal energies the stress-bearing network within the gel restructures over time as individual particle bonds break and reform. Typically, under gravity such weak gels show a prolonged period of either no or very slow settling, followed by a sudden and rapid collapse - a phenomenon known as delayed collapse. The link between local bond breaking events and the macroscopic process of delayed collapse is not well understood. Here we summarize the main features of delayed collapse and discuss the microscopic processes which cause it. We present a plausible model which connects the kinetics of bond breaking to gel collapse and test the model by exploring the effect of an applied external force on the stability of a gel.

\end{abstract}

\maketitle
 \makeatletter
    \providecommand\@dotsep{5}
  \makeatother
%  \listoftodos\relax

%\linenumbers

%
%%====================================================================
\section{Introduction}
\label{sec:introduction}
%%====================================================================

%TODO: Nomenclature

\nomenclature[\a]{$\a$}{Radius of particle \verb+\a+}
\nomenclature[\cpstar]{$\cpstar$}{Scaled polymer concentration \verb+\cpstar+}
\nomenclature[B]{$B$}{ratio of buoyant stress to compressive yield stress of gel \verb+B+}
\nomenclature[\D]{$\D$}{range of attractive potential \verb+\D+}
\nomenclature[\delta rho]{$\Delta \rho$}{Density difference \verb+\Delta \rho+}
\nomenclature[\etaL]{$\etaL$}{low-shear limiting viscosity \verb+\etaL \rho+}
\nomenclature[g]{$g$}{Acceleration due to gravity \verb+g+}
\nomenclature[h]{$h$}{height of gel \verb+h+}
\nomenclature[\hcrit]{$\hcrit$}{critical height of gel \verb+\hcrit+}
\nomenclature[\h0]{$\h0$}{Initial height of gel \verb+\h0+}
\nomenclature[\k0]{$\k0$}{Initial permeability of gel \verb+\k0+}
\nomenclature[\lcrit]{$\lcrit$}{stress transmission length \verb+\lcrit+}
\nomenclature[\lg]{$\lg$}{mean width of strands \verb+\lg+}
\nomenclature[M_{\mys{w}}]{$M_{\mys{w}}$}{Polymer molecular weight \verb+M_{\mys{w}}+}
\nomenclature[\NA]{$\NA$}{Avocadro's constant \verb+\NA+}
\nomenclature[\phicolloid]{$\phicolloid$}{Colloid volume fraction \verb+\phicolloid+}
\nomenclature[\phic]{$\phic$}{Initial colloid volume fraction \verb+\phic+}
\nomenclature[\rg]{$\rg$}{Polymer radius of gyration \verb+\rg+}
\nomenclature[\sigcrit]{$\sigcrit$}{Critical stress for solid-liquid transiton \verb+\sigcrit+}
\nomenclature[\sigg]{$\sigg$}{Buoyant stress on gel \verb+\sigg+}
\nomenclature[\sigyield]{$\sigyield$}{compressive yield stress \verb+\sigyield+}
\nomenclature[\sigma]{$\sigma$}{Stress variable \verb+\sigma+}
\nomenclature[\taulife]{$\taulife$}{lifetime of percolating network of gel \verb+\taulife+}
\nomenclature[\tausec]{$\tauesc$}{Kramers escape time \verb+\tauesc+}
\nomenclature[\taubreak]{$\taubreak$}{Stress dependent time for solid-liquid transiton \verb+\taubreak+}
\nomenclature[\taumean]{$\taumean$}{Mean delay time}
\nomenclature[\tau0]{$\tau0$}{Characteristic diffusion time}
\nomenclature[t]{$t$}{general time variable \verb+t+}
\nomenclature[\tw]{$\tw$}{Waiting time \verb+\tw+}
\nomenclature[\Uc]{$\Uc$}{interparticle attractive energy at contact \verb+\Uc+}
\nomenclature[\v0]{$\v0$}{settling velocity of particle at infinite dilution \verb+\v0+}
\nomenclature[\vs]{$\vs$}{Slow settling rate \verb+\vs+}
\nomenclature[\vf]{$\vf$}{Fast settling rate \verb+\vf+}
\nomenclature[\xi]{$\xi$}{hydrodynamic resistance due to flow of solvent in gel \verb+\xi+}

Colloidal gels are components of everyday products such as foodstuffs, fabric conditioners, cosmetics, shampoos, and even toothpaste, yet despite their practical importance they present many challenges to our understanding of disordered materials. A gel is a solid containing a network of particles which is formed when a colloidal dispersion with attractive interactions  $\Uc$  is quenched deep into a two-phase region of phase space \cite{7444}. Driven far out-of-equilibrium, the kinetics of phase separation are dramatically slowed down or, in the limit of strong short-range attractions ($\D / \a \ll 0.1$ with $\D$  the range of the attractive interactions and $a$ the particle radius), totally arrested. Partial phase separation generates a disordered network, whose initial structure is controlled by the strength of interaction $\Uc$, the range $\D / \a$ of the potential, as well as the volume fraction $\phicolloid$ of colloids. The long-time structural integrity of this network is, in the majority of cases, constrained by the gravitational stress exerted by its own weight.  Given sufficient time, a gel settles under gravity, if it is not density-matched.  We distinguish two limiting cases: In strong gels, where the attractive interactions are large in magnitude ($\Uc  \gtrsim 20 \kBT $ and narrow in range $\D / \a \ll 0.1$), compaction occurs smoothly at a rate which decreases progressively with time. The time dependence of the height of the gel in this case is well described by a poroelastic settling model \cite{5922,3708,11697}; By contrast, in weak gels where attractions are comparable to $\kBT$ and wide in range $\D / \a > 0.1$ so that thermal fluctuations are significant, an anomalous behaviour called {\it delayed collapse} is observed.  A weak gel, instead of instantaneously settling, hesitates for a well defined delay period $\taud$ without any sign of macroscopic settling, before suddenly undergoing a rapid and catastrophic collapse. Delayed collapse has been observed in a wide variety of systems \cite{5744,5926,Allain-1364,5382,2280,2308,2905,5390,5389,5237,5241,10832,12095} so the response appears to be a universal feature of weak gels yet to-date no theoretical framework has emerged to account for this process. 

The existence of a measurable delay in the collapse of a weak colloidal gel looks, at first sight, rather surprising. In a crystalline solid yield usually happens spontaneously at a well defined stress and there is no latency before the material flows. Similarly, if a constant gravitational stress is applied to a free-flowing colloidal suspension sedimentation  occurs essentially instantaneously. The anomalous response evident in a weak gel has been interpreted as the signature of a non-equi\-librium solid-to-fluid transition, triggered by erosion of the gel network by internal flows \cite{5744,Allain-1364,2530,2308,5390,9091}, progressive fracture by an applied gravitation stress \cite{2280}, or as a consequence of thermal restructuring of the network \cite{5382,5237,12095}.  The origins of delayed collapse are both  scientifically  fascinating as well as being technologically relevant because colloidal gels are often used  to stabilize complex product formulations against macroscopic phase separation. The network of particles supports the gravitational stress exerted by the formulation and suppresses unwanted sedimentation of the product. This trick, widely used by formulators, works only for $\tw < \taud$ (where $\tw$ is the age of the gel) as the gel instability at $\taud$ eventually restores equilibrium and phase separation starts again. In such cases, $\taud$ fixes the ultimate physical self-life of the product. A demand for robust long-life formulations has heightened the need for a better microscopic understanding of the process of delayed collapse so that gel instability may be predicted and controlled. 

Much of the  work to date on gel collapse has focused on macroscopic features, typically by measuring the time evolution of the height $h(\tw)$ of a gel, rather than on the internal structure and dynamics of a gel. However in the last few years new techniques, such as confocal scanning microscopy \cite{5241,10832,12095} and photon correlation imaging \cite{11697} has revealed that colloidal gels have a complex hierarchical structure, with different structural features at different length scales. So while at the individual particle level, a gel consists of dense aggregated colloids, the aggregates are organized on the micro-scale into relatively thick strands of particles, which at the mesoscale  are assembled into a percolating network able to transmit a stress. In a weak gel, the stress bearing network  has a number of distinctive characteristics. First, it is mechanically heterogeneous with a complex structure consisting of weakly connected soft regions of low particle density coexisting with stiff dense strands of spheres. The strands of the gel may be several radii thick, depending on the strength and range of the attractive interactions. Second, the network is dynamic and restructures with time, as inter-particle bonds break and reform via thermal fluctuations. A microscopic understanding of just how such a spatially and temporally disordered network evolves in time and how it transmits stress is still elusive yet just such an insight is essential for the prediction and control of delayed collapse.

The goal of this paper is to summarize the key features of delayed collapse in weak depletion gels, to speculate on their origins, and to identify the key issues that still remain to be resolved. We briefly review the features of delayed collapse seen in the gravitational settling of weak gels. With a few recent exceptions, the majority of the previous work reported to date has focused on macroscopic features such as the time evolution of the height of a gel. We discuss the microscopic changes in the gel as it restructures by the breaking and reforming of inter-particle bonds. Using this experimental insight, we propose a new model for delayed collapse which emphasises the connectivity of the stress-transmitting network. Finally, we explore the role of an external applied force on the delay time of a gel and interpret the results in light of the new model.

%
%%====================================================================
\section{Delayed collapse}
\label{sec:collapse}
%%====================================================================

%\todo[inline]{Figure captions}

\begin{figure}[ht]
		\includegraphics[width=0.9\textwidth]{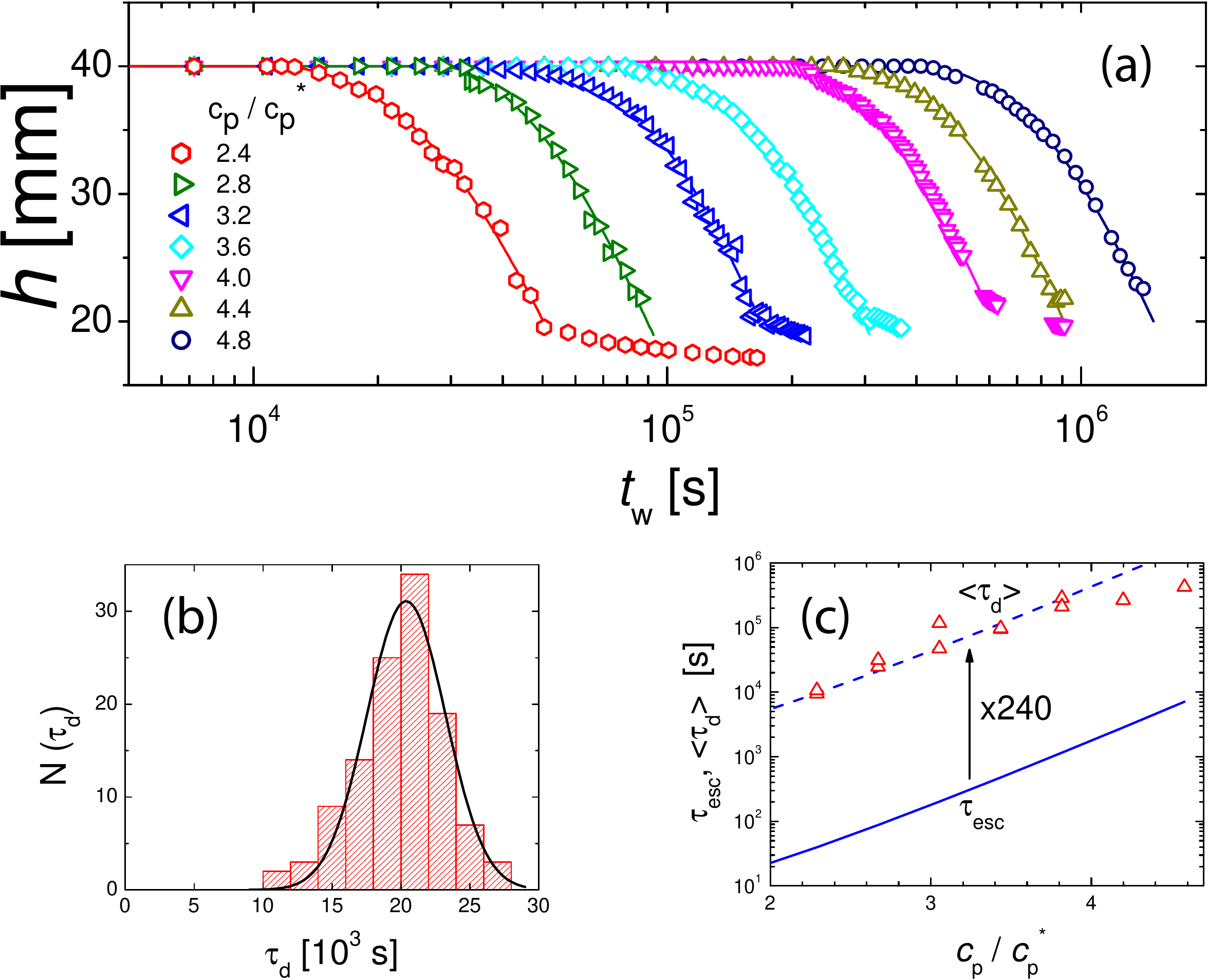}
        \caption{(color online). Delayed collapse in a colloidal gel. (a) Time evolution of the total height $h(\tw)$ of a depletion gel  ($\Delta / \a = 0.62$, $\a = 316$ nm)  with $\phicolloid = 0.21$, and different polymer concentration $\cpstar$. (b) Probability distribution $N(\taud)$ for the delay time of a gel  with $\Delta / \a = 0.45$, $\a = 272$ nm at $\phicolloid = 0.20$, $c_{\mys{p}} / c_{\mys{crit}} = 4$. The solid line shows a fit to a Gaussian. (c) Dependence of the mean delay time $\taumean$ on the polymer concentrations for a gel with $\Delta / \a = 0.62$, $\a =316$ nm, $\phicolloid = 0.21$.  The solid line depicts the estimated single bond lifetime $\tauesc$. }
        \label{fig:mockup}
\end{figure}

In this section, we recall the principle \textit{macroscopic} features of delayed collapse in weak gels. We focus on the time-dependent evolution of the total height $h(\tw)$ of a gel in a gravitational field, as a function of its age $\tw$. Typically a suspension is randomized by shaking or mixing at time $\tw = 0$ and then left undisturbed during the sedimentation process. The height $h(\tw)$ usually displays three regimes \cite{12095}: an initial lag period of width $\taud$ where the height falls slowly but continuously with age, a non-linear regime of rapid collapse where the interface velocity $\mathrm{d}h / \mathrm{d} \tw$ speeds up with $\tw$, and finally a region of compaction where the height relaxes asymptotically towards an equilibrium value. The three-stage nature of delayed collapse is exemplified by the data reproduced in Fig.~\ref{fig:mockup}(a) on the settling of depletion gels \cite{12095} with $\phicolloid = 0.21$ for polymer concentrations $\cpstar  \in [2.4,4.8]$. We use two sizes of low polydispersity surfactant-stabilized poly(dimethyl siloxane) (PDMS) emulsion droplets suspended in a refractive-index and near-density matched mixture of 1,2-ethane diol and water. The suspension of large droplets had a hydrodynamic radius of $\a =316 \pm 11$ nm and a polydispersity of 0.17 while the small droplets had a radius of  $\a =272 \pm 10$ nm and a polydispersity of 0.18. A depletion attraction was induced by the addition of either the anionic polyelectrolyte xanthan ($M_{\mys{w}}$ = 4.66 x 10$^{6}$ g mol$^{-1}$, radius of gyration $\rg = 194$ nm) or neutral hydroxyethylcellulose ($M_{\mys{w}}$ = 1.3 x 10$^{6}$ g mol$^{-1}$, radius of gyration $\rg = 126$ nm) depending on the range of attractions required. The majority of experiments were conducted using a combination of large emulsion droplets and xanthan to give a colloid-polymer mixture with an attractive range of $\rg / \a = 0.62 \pm 0.04$ \cite{11551}. A small number of results were obtained using  a colloid-polymer mixture with an attractive range of $\rg / \a = 0.45 $ \cite{Zhang2013}, obtained by combining small emulsion droplets with hydroxyethylcellulose. The magnitude of the attractions between droplets was adjusted by varying the concentration $\cpstar$ of added polymer, which is expressed in units of the overlap concentration $\cp^{*}$. The relative buoyancy of the emulsion droplets is $\Delta \rho =  \rho_{\mys{c}} - \rho_{\mys{s}} = -130$ kg m$^{-3}$, with $\rho_{\mys{c}}$ and $\rho_{\mys{s}}$ the densities of the droplet and solvent mixture respectively. This relatively small density difference explains the absence, evident in Fig.~\ref{fig:mockup}(a),  of any discernible settling during the lag phase. This is in contrast to many of the earlier studies where the initial stages of gel settling were often characterized by a relatively broad change in the height around $\tw = \taud$ as the gel was already settling slowly before the regime of rapid collapse started. This makes it more difficult to identify precisely the point where rapid collapse starts.  In the PDMS system it is straight-forward to determine the characteristic delay time $\taud$, which is taken as the time at which the height $h(\tw)$ of the gel first begins to noticeably drop from its initial value $\h0$. In this paper we concentrate on the physical and chemical factors that determine the duration $\taud$ of this initial lag period.

% corresponds to a gravitational length $l = \kBT /(\frac{4}{3} \pi \a^{3} |\Delta \rho| g)$ of $\approx 77$ radii so that the gravitational field provides only a weak perturbation on the microscopic structure of the gel. \note{1[1]} The small value of $|\Delta \rho|$ also 

Experiments reveal \cite{11551,12095} that the characteristic delay time $\taud$ has a number of distinctive characteristics: First, measurements of $\taud$ from samples of identical materials display a significant statistical variance \cite{2308}; and second, the average delay time  $\taumean$ is very sensitive to the magnitude of the attractive interactions {\it i.e.} doubling the polymer concentration in a depletion system may increase the delay time by 1--2 orders of magnitude \cite{5926,2905,5237,9091,10832,12095}. Figures~\ref{fig:mockup}(b) and (c) confirms the generality of these conclusions in the PDMS system. To establish the extent of statistical variations in the delay time, we performed 116 repeat measurements of the delay time on a depletion gel \cite{Zhang2013} with $\Delta / \a = 0.45$ ($\a = 272$ nm) in a temperature controlled environment with $\Delta T = 24.7 \pm 0.1 \degc$. Each of the runs was made on a freshly-prepared gel sample under identical conditions to ensure that individual runs were uncorrelated. The resulting values of $\taud$ were used to construct the probability distribution $N(\taud)$ of delay times shown in Fig.~\ref{fig:mockup}(b). An appreciable variation in the measured delay times is seen with a scatter of about 14 \%. The distribution $N(\taud)$ is symmetric and well fitted by a Gaussian distribution (solid line in Fig.~\ref{fig:mockup}(b)) which suggests that delayed collapse is a consequence of a large number of independent uncorrelated stochastic events. 

%The nature of the stochastic events responsible for delayed collapse is hinted at 

A clue to the nature of the stochastic events responsible for delayed collapse is revealed by the strong correlation evident in Fig.~\ref{fig:mockup}(c) between the mean delay time $\taumean$ and the  lifetime $\tauesc$ of an individual colloid-colloid bond. To estimate the bond lifetime we assume that the rupture of individual bonds occurs primarily as a consequence of thermal fluctuations and that the gravitational stress (which as  noted above is small in the PDMS system) does not significantly accelerate the rate of bond rupture. When no force is applied across the bond, the lifetime $\tauesc$ may then be expressed in terms of the average time it takes a Brownian particle to diffuse out of the attractive potential (the Kramers' escape time). In the limit of  $\Uc \gg \kBT$, the lifetime is given by the Arrhenius expression, $\tauesc =  \tauzero \exp (\Uc / \kBT)$, where $\tauzero$ is a characteristic time which depends on the colloid diffusivity and the range and depth of the interaction potential. Estimating $\tauzero$ from the measured low shear rheology, and the width $\Delta$ and depth $\Uc$ of the depletion potential from accurate generalised free-volume theories \cite{12095} yields the lifetimes plotted in Fig.~\ref{fig:mockup}(c). As may be seen, the ratio of the two characteristic times is very nearly constant, for a wide range of polymer concentrations, with the delay time approximately 240 times the bond lifetime.  The strong correlation evident in Fig.~\ref{fig:mockup}(c) highlights the pivotal role of spontaneous thermal fluctuations. The observation that delayed collapse occurs on timescales two orders of magnitude larger than the bond rupture time indicates that a microscopic particle-level model is inadequate to account for collapse. The large difference between $\taumean$ and $\tauesc$ highlights the crucial role played by the hierarchical structure of a gel. The characteristic ratio  $\taumean / \tauesc$ is also likely to depend on the initial volume fraction $\phi$ of the gel but this dependence has, to date, been little studied experimentally.

The delay time, in addition to its sensitivity to the interaction potential and the particle volume fraction $\phi$, has also been reported to depend on physical factors  such as the height, width, and the cross-sectional shape of the sample cell \cite{5744,2308}. Starrs\emph{ et al.}, for instance, observed  that weak depletion gels constructed from poly(methyl methacrylate) spheres with initial he\-ights $\h0$ above  7 mm, showed a height-independent delay time \cite{2308}. However in one sample with $\h0 = 5$ mm there was a marked increase in the delay time  by a factor of about 70\%. Evans and Starrs \cite{2530} interpreted these observations as evidence for the existence in a gel of a new macroscopic length scale $\lcrit$, which they termed a `stress transmission length'.  They argued that samples whose height $\h0$ were greater than $\lcrit$ could not transmit gravitational stress to the base of a sample so no properties of the gel could be a function of $\h0$. Aside from this study, the effect of height on gel collapse has not been systematically studied although Kim \emph{et al.} \cite{5243} and others have noted a qualitative change in settling with short samples generally showing a steady or `creeping sedimentation' while delayed collapse seemed only to be shown by tall samples. Interpretation of these observations is complicated by the relatively small range of initial heights used, but suggest that collapse phenomena might show a rich dependence on the initial height. 

\begin{figure}[ht]
        \begin{center}
        \includegraphics[width=0.7\textwidth]{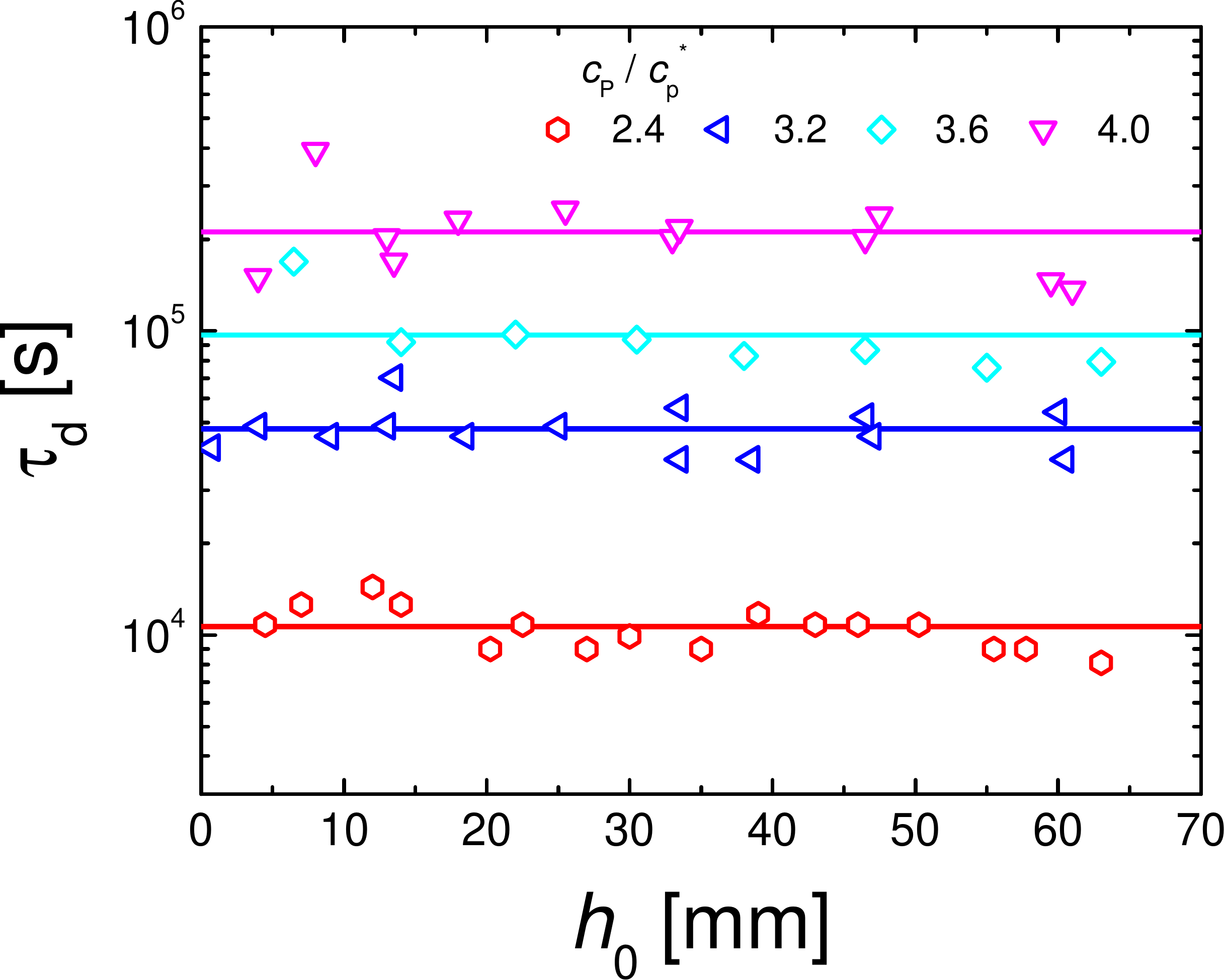}
        \caption{(color online). Dependence of the delay time on the initial height of a gel. The data is for a depletion gel with $\Delta /\a = 0.62$, $\a =316$ nm,  $\phicolloid = 0.21$ and four different polymer concentrations.}
        \label{fig:height-delay}
        \end{center}

\end{figure}

To clarify the effect of height on collapse, we used two complementary measurement techniques to probe the delay process in the PDMS system over a wide range of initial heights. For relatively tall samples ($\h0 \gtrsim 5$ mm) a CCD camera was used to measure accurately the time dependence of the height, $h(\tw)$, of the gel with a spatial resolution of about 0.5 mm.  In short samples the collapse kinetics was followed by confocal scanning laser microscopy. The gels were repeatedly imaged by scanning a large number ($\sim 100$) of slices perpendicular to the gravitational field at different time intervals (from minutes to days). The fluorescent intensity in each slice was integrated to give a height profile. Both the confocal and CCD experiments were performed in cylindrical glass cells with a constant width of 17 mm to eliminate the effects of width and cell geometry.  The delay time, $\taud$, for gels with different polymer concentrations $\cpstar$ is shown in Fig.~\ref{fig:height-delay} as a function of the initial height, $\h0$. Fig.~\ref{fig:height-delay} indicates convincingly, at least for this system, that the time lag before the initiation of the rapid collapse process is independent of the initial height of the sample.  These observations confirm the conclusions of the more limited set of height-data ($\h0 > 20$ mm) reported in \cite{12095}. The independence of $\taud$ on $\h0$  is particularly striking because the initial height of the gel was varied  over almost two orders in magnitude, from 0.78 mm to 63 mm. This behaviour is in contrast with the more limited data reported by Starrs \textit{et al} \cite{2308} who observed a height-dependent delay. More experiments will clearly be needed to fully elucidate the general picture but the data in Fig.~\ref{fig:height-delay} suggests that a dependence on height at least is probably not a universal feature of delayed collapse.

%%====================================================================
\section{Microscopic picture of gel restructuring}
\label{sec:microstructure}
%%====================================================================

\begin{figure}[tbp]
    \begin{center}
    \includegraphics[width=0.6\textwidth]{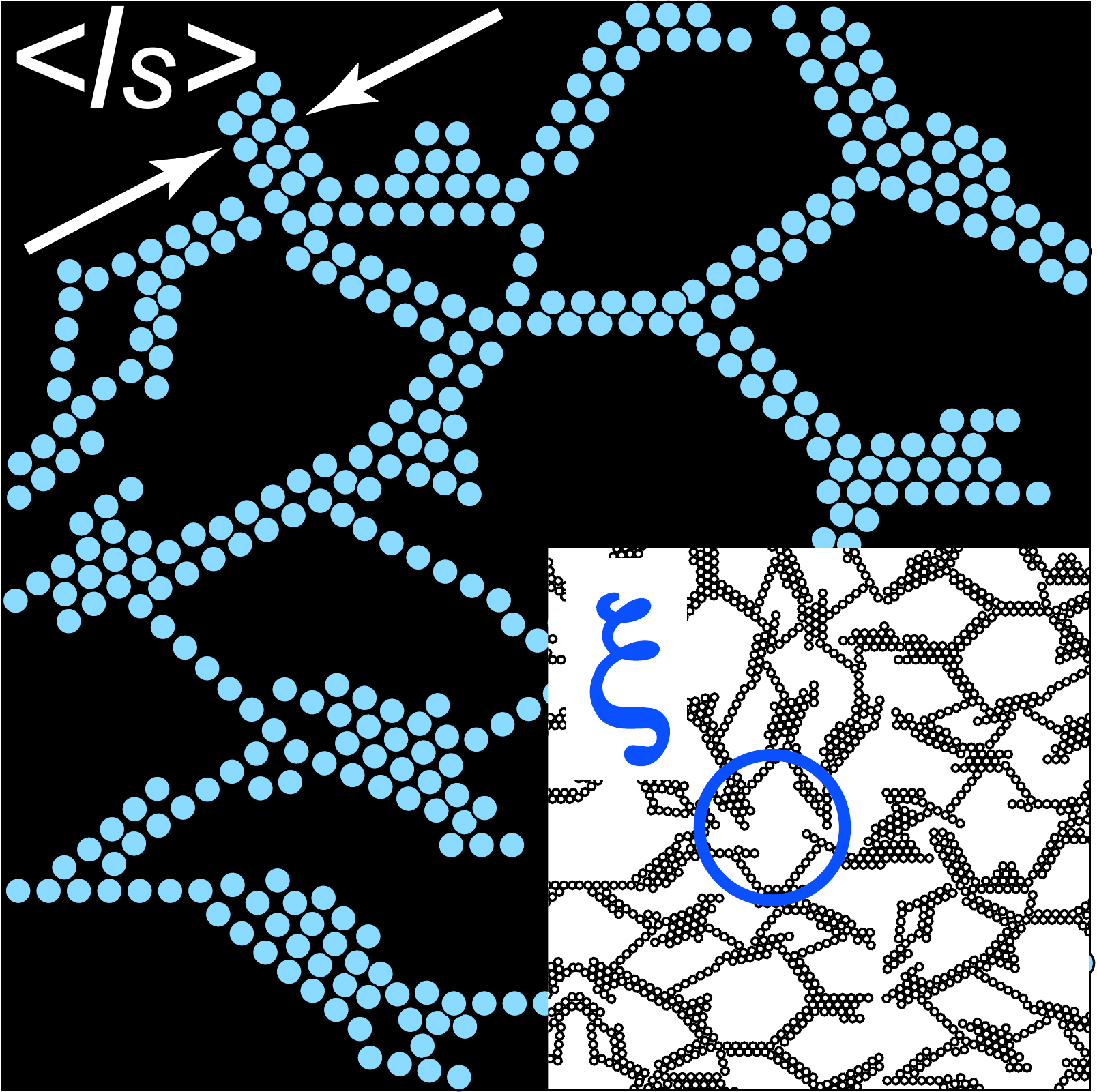}
    \caption{(color online). Schematic of a gel illustrating the two length scales, $\lg$ and $\R$, which characterize the disordered structure. The  mean chord length $\lg$ is a measure of the average diameter of the particle strands while the correlation length $\R$ is a typical size of the pores within the framework of the gel.}%
     \label{fig:gel}
     \end{center}
\end{figure}

During the delay period, although there is no loss of overall mechanical integrity, the gel continuously restructures as thermal fluctuations favour the breaking of existing particle bonds and the formation of new ones. To understand the consequences of thermal rearrangements for the hierarchical structure of a gel we have used fluorescent confocal microscopy on a labelled PDMS gel to directly follow the time evolution of the network in real space \cite{12095}. %Figure~\ref{fgr:ageing}(a) shows an example of the stress-bearing network found. 
We characterize how the micro-structure changes by determining two characteristic length scale, the correlation length $\R$ and the mean chord length $\lg$, which are illustrated in Fig.~\ref{fig:gel}. A series of two-dimensional confocal images  were recorded, at a fixed height within the gel, as a function of the sample age $\tw$. To characterise the large-scale fluctuations in the particle density seen in the confocal images the structure factor $S(q)$ was calculated from a 2D Fourier transform and the characteristic wave-number of the intensity peak evaluated as $q = \int \mathrm{d}q qS(q) / \int \mathrm{d}q S(q)$. The existence of a peak in the scattering intensity $S(q)$ at small scattering vectors is a ubiquitous feature of colloidal aggregation \cite{Cipelletti-1517}.  The peak signifies the presence of a correlation length $\R \sim \pi / q$. In the dilute particle limit $\phicolloid \rightarrow 0$, this characteristic length scale arises from the presence of close-packed fractal clusters of size  $\R$, while at higher concentrations it may be more accurately interpreted as a measure of the heterogeneity or mesh size of the gel.   To quantify changes in the \textit{local} micro-structure the mean chord (or intercept) length $\lg$ was measured. $\lg$ is the average extent of a randomly-orientated line segment which lies totally within the particle rich portion of the gel. Since the average is dominated by orientations perpendicular to the strands of particles $\lg$ is a convenient experimental measure of the strand diameter. 

\begin{figure}[h]
\centering
  \includegraphics[width=1.0\textwidth]{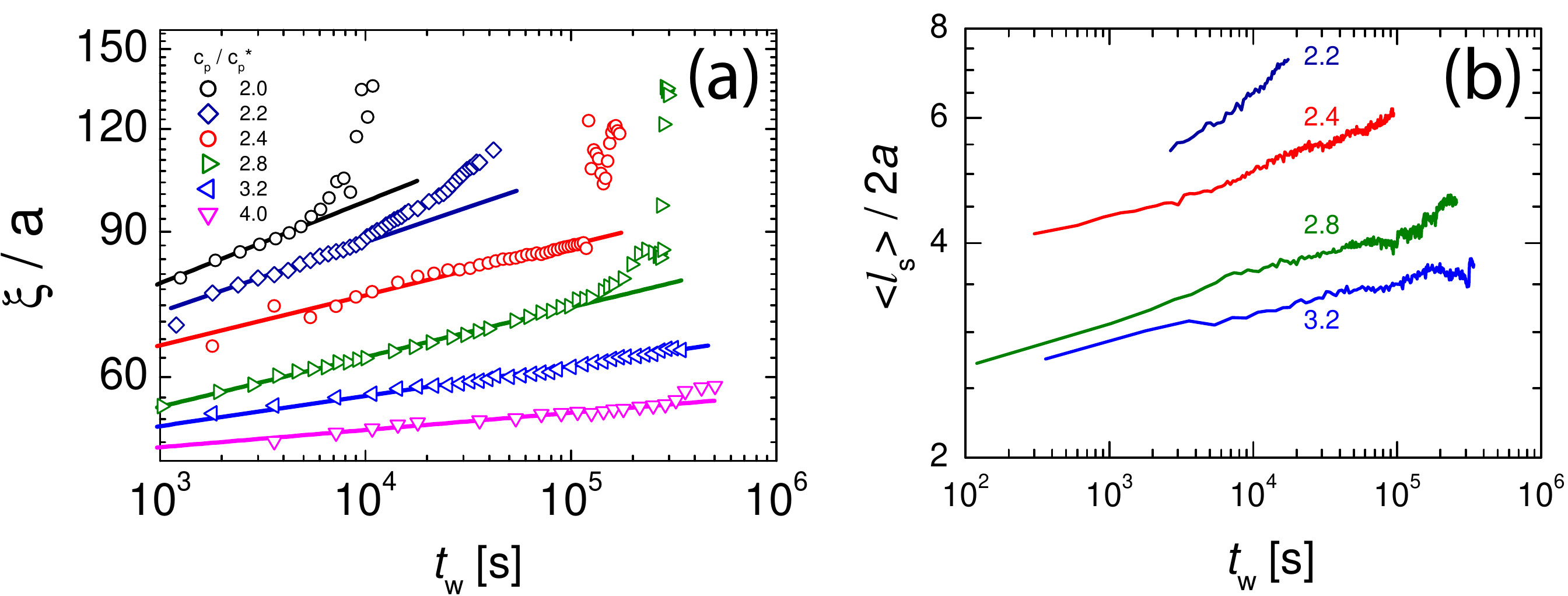}
  \caption{(color online). Restructuring of a thermal gel ($\Delta /\a = 0.62$, $\a = 316$ nm,  $\phicolloid = 0.21$). Time evolution of (a) the correlation length $\R$, and (b) the mean strand thickness $\lg$, for indicated polymer concentrations.}
  \label{fgr:ageing}
\end{figure}

The interparticle processes which occur as a gel ages are illustrated by the confocal measurements, summarized in Figures~\ref{fgr:ageing}(a) and (b). The PDMS gel has a hierarchical structure with two clearly separable length scales. The confocal images reveals that at the microscale, the attractive particles aggregate into relatively thick gel strands which, for the PDMS gel, are between 3 and 8 diameters in width. The widths of the strands are observed to depend on the age of the gel and the strength of the attractive potential. At the next level of organization, the strands form on the mesoscale a percolating network with a correlation length $\R$ which ranges from between 50 and 100 particle radii in extent for the PDMS system. Both the correlation length $\R$ and the strand thickness $\lg$  grow continuously with the age $\tw$ of the gel and show no sign of the dynamical arrest transition expected for a strong gel. The rate of coarsening of the gel network is observed to depend sensitively on the strength of the attractive interactions $\Uc$. The more attractive the interparticle potential at contact, the slower is the observed rate of coarsening \cite{11551}.

\begin{figure}[tbp]
    \begin{center}
    \includegraphics[width=0.7\textwidth]{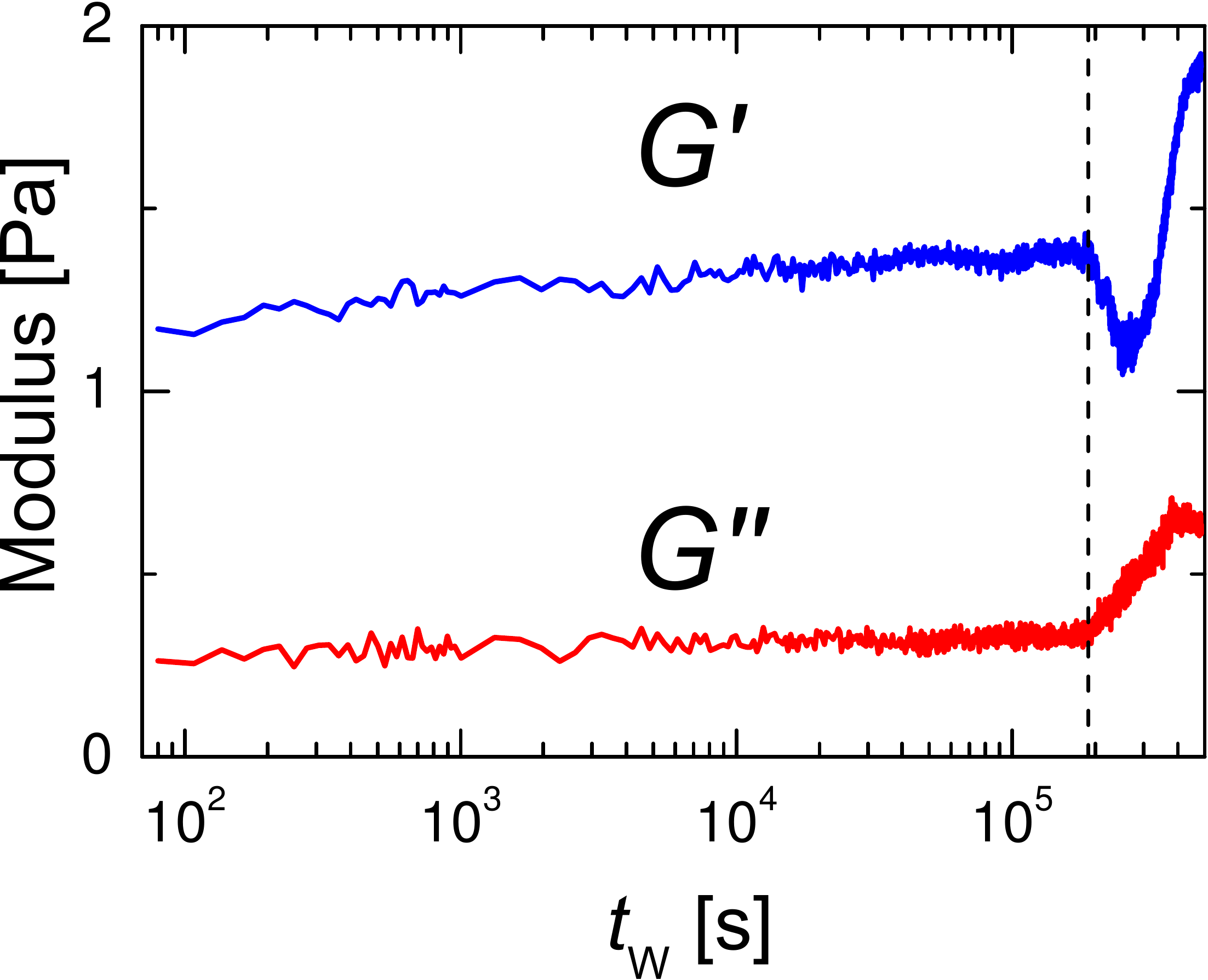}
    \caption{(color online). Elastic $G'$ and loss $G''$ moduli of a weak gel ($\Delta /\a = 0.62$, $\a = 316$ nm,  $\phicolloid = 0.21$, $\cpstar = 4.0$) as a function of time elapsed since preparation $\tw$. The mechanical properties were measured by applying an oscillatory stress of 0.0025 Pa at a frequency of 0.5 Hz. }%
     \label{fig:rheology}
     \end{center}
\end{figure}

The picture which emerges of a network which grows progressively coarser and thicker with time implies that the shear elasticity $G'$ should also increase with $\tw$. This conjecture is confirmed by rheological measurements shown in Figure~\ref{fig:rheology} where  $G'$ increases with $\tw$, prior to collapse. This however leads to an apparent contradiction -- why if the gel network is actually getting stiffer with time, does it ever collapse under a gravitational stress?  One possibility is that it is not the average micro-structure which is important but the connectivity of the gel, and in particular how the ability of the network to transmit stress evolves with time. In this scenario, collapse occurs as ageing reduces the connectivity and increases the fragility of the network. In a network with few connections a bond breaking event results in large scale cooperative displacements, far away from the event that caused it. The response is highly non-local. A relatively small number of independent bond-breaking events, distributed randomly throughout such a fragile network, would then trigger a  catastrophic macroscopic failure of the gel.

\begin{figure}[tbp]
    \begin{center}
    \includegraphics[width=0.65\textwidth]{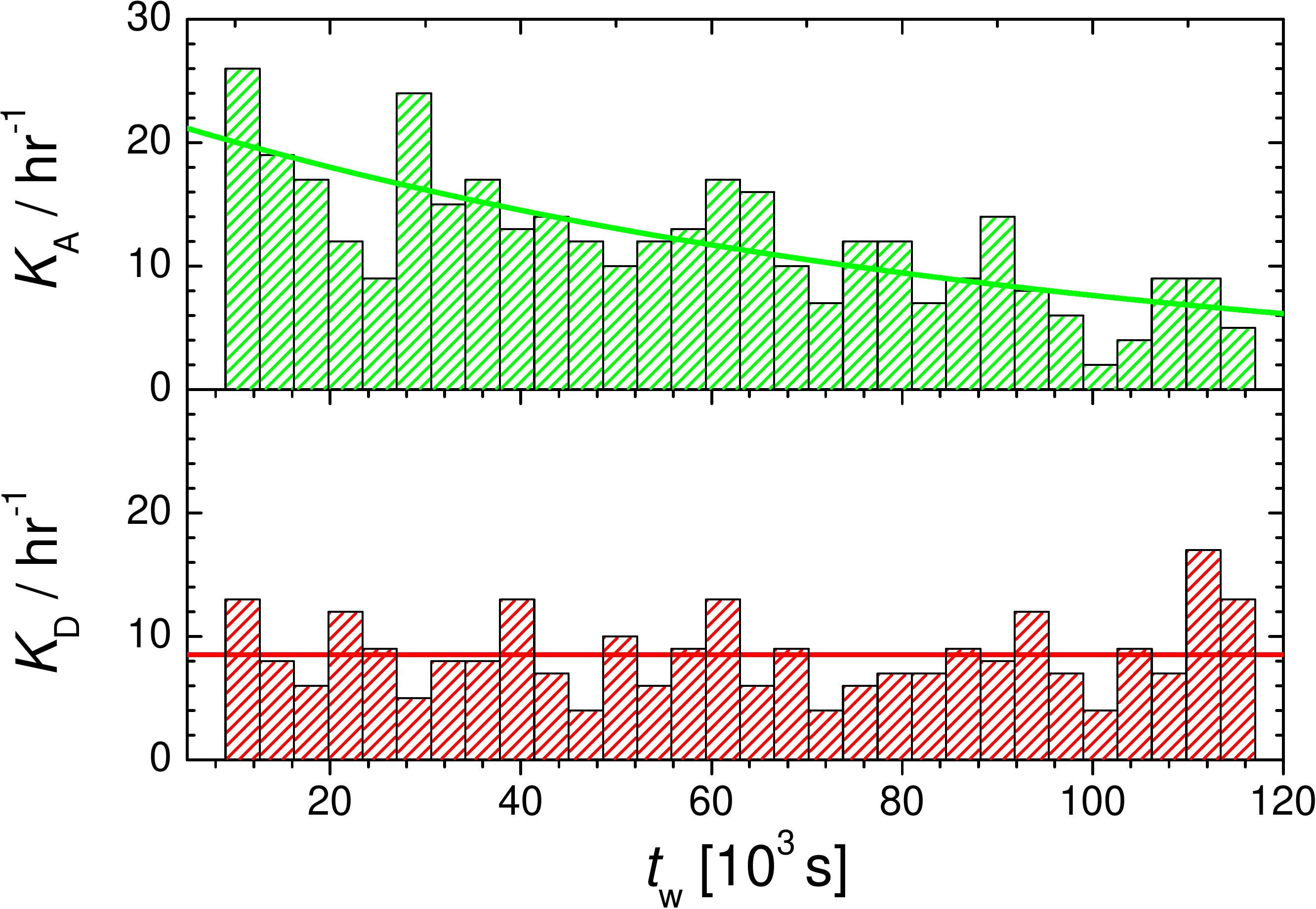}
    \caption{(color online). The number of association $K_{\mys{A}}$ and dissociation $K_{\mys{D}}$ events per hour as a function of the age $\tw$ of a restructuring gel ($\Delta /\a = 0.62$, $\a = 316$ nm, $\phicolloid = 0.21$, $\cpstar = 2.4$). The lines are guides to the eye.  Delayed collapse occurs at $\taud \sim 120 \times 10^{3}$ s.}%
     \label{fig:aging-events}
     \end{center}
\end{figure}

Support for the importance of connectivity comes from confocal microscope studies \cite{12095}. A relatively large random region within a PDMS gel was imaged hourly, for a total of 32 hours. Comparing consecutive images,  the number of discrete strand association and dissociation events occurring per hour were identified. The results for the corresponding rates for strand association  $K_{\mys{A}}$, and strand dissociation  $K_{\mys{D}}$   are plotted in Figure~\ref{fig:aging-events}, as a function of the age $\tw$ of the gel. The data reveals a number of distinctive features. First, while the dissociation rate seems to be essentially unaffected by the age of the gel, the rate of association drops rapidly with increasing $\tw$.  The two time dependences reflect the radically different physical origins of the processes involved. Dissociation is an activated process which requires particles to escape from a deep potential well  and so is not expected to depend on the age of the sample. While, the decline in the rate of association with $\tw$ reflects the progressive slowing down of the microscopic dynamics frequently seen in soft glassy systems as they evolve towards a more homogeneous state \cite{4213}. The second distinctive feature of Figure~\ref{fig:aging-events} is the dominance at early times of association over dissociation events. The data reveals that dissociation is a relatively rare event and a strand once broken will be almost immediately reformed by an association event so that the connectivity of the network changes only very slowly with $\tw$. The network is essentially self-healing and reforms rapidly after any deformation. At longer times, the character of the network changes as the rates of association and dissociation become comparable. At this point the connectivity of the network  drops rapidly with increasing age. The degradation of the network continues to a point at which a small number of uncorrelated dissociation events result in a rapid macroscopic failure of the gel. In support of these ideas, we find that the time at which $K_{\mys{A}} \approx K_{\mys{D}}$ correlates well with the macroscopic delay time $\taud$.

%%====================================================================
\section{Microscopic model of collapse}
\label{sec:model}
%%====================================================================

We now propose a microscopic model which connects the kinetics of bond-breaking to gel collapse and use it to predict the effect of a small applied force on the time $\taud$ for gravitational failure. Our approach is based on the recent work of Lindstr\"{o}m et al. \cite{lindstrom_structures_2012} but modified to analyse a weak thermal-fluctuating gel. 

We begin by proposing that the delay time $\taud$ is determined by the time-scale at which the rate of strand association becomes comparable to the rate of strand dissociation. On this time-scale the subsequent lose of connectivity of the network and the eventual failure of the gel is rapid  by comparison, and so can safely be neglected. Motivated by the experimental data presented above, we assume that the rate $K_{\mys{D}}$ at which a gel strand breaks is time-independent while $K_{\mys{A}}$, the rate at which the strands of the network associate and re-form, depends on the age of the gel.  While the quench history of the gel, and its chemistry will,  in general have a significant impact on the time dependence of $K_{\mys{A}}(\tw)$,  close to the vicinity of the collapse point the data in Fig.~\ref{fig:aging-events} suggests  the time dependence of $K_{\mys{A}}(\tw)$ may be approximated by a simple linear function of gradient $\mathrm{d} K_{\mys{A}}(\tw) / \mathrm{d}\tw = -\alpha$, where $\alpha$ is a positive constant.

When a small perturbative force $f$ (which is defined here to be compressive if $f > 0$) is applied across a weak attractive particle-particle bond, the activation energy for bond dissociation is increased by a factor of $f \cdotp \Delta$ where $\Delta$ is the width of the attractive potential \cite{2161}, as illustrated in Fig.~\ref{fig:potential}(a). The corresponding bond dissociation rate changes to
\begin{equation}
  k =  \frac{1}{\tauesc} \exp \left (   -\frac{f} {f_{\mathrm{th}}}  \right )
\end{equation}
where $\tauesc$ is the bond lifetime when no force is present and $f_{\mathrm{th}}= \kBT/\Delta$ is the characteristic force for thermal bond rupture. The effect of a weak point force on the gel is more complex however because the failure of the gel is dictated by the rate of strand dissociation $K_\mys{D}$, which because a strand is several particles thick, could be significantly slower than the dissociation of an individual particle bond. Lindstr\"{o}m et al. \cite{lindstrom_structures_2012} have derived expressions for the cooperative dissociation rate of a strand of $n$ particles in different dynamical regimes when either association or dissociation events dominate. The experimental data in Fig.~\ref{fig:aging-events} suggests that,  in our system, the number of association events is roughly comparable to the number of dissociation events with the ratio $K_\mys{A} / K_\mys{D}$ becoming no larger than about two even for $\tw \rightarrow 0$. In this case, the analysis \cite{lindstrom_structures_2012} of Lindstr\"{o}m et al. suggests that the dissociation rate of a strand with many bonds in its cross-section should be almost the same as the dissociation rate of the individual bonds that make up the strand, so that
\begin{equation}\label{eq:strand}
K_\mys{D}(f) = K_\mys{D}(0) \cdotp \exp \left (  - \frac{f} {f_{\mathrm{th}}}  \right )
\end{equation}
where $K_\mys{D}(0)$ is the dissociation rate of the strand in the absence of any applied force. 

Within this model, applying a counteracting force to the gel will reduce the rate at which the strands of the network breakup, postponing the approach to the instability condition  $K_\mys{A} \approx K_\mys{D}$, and consequently increasing the delay time $\taud$. If a force $f$ is applied across each particle bond, collapse occurs after an interval $\taud$, which is the solution of the implicit equation $K_{\mys{A}}(\tw = \taud) =  K_{\mys{D}}(f)$. To solve this equation, we recognize that Fig.~\ref{fig:aging-events} suggests that the dissociation rate is time-independent while the association rate varies approximately linearly with the elapsed time,
\begin{equation}
K_\mys{A}(\tw) = K_\mys{A}^{0} - \alpha \tw.
\end{equation}
Here $K_\mys{A}^{0}$ is the initial ($\tw \rightarrow 0$) rate of association.  An expression for the dependence of the delay time $\taud(f)$ on the applied force $f$,
\begin{equation}\label{eqn:model}
\frac{\taud(f)}{ \taud(0)} = 1 + \frac{K_\mys{D}(0)}{K_\mys{A}^{0} - K_\mys{D}(0)} \left [ 1 - \exp \left (  - \frac{f} {f_{\mathrm{th}}}  \right ) \right ],
\end{equation}
follows directly from the collapse condition, $K_{\mys{A}}(\taud) =  K_{\mys{D}}(f)$, and Eq.~\ref{eq:strand}. Here $\taud(0) = (K_\mys{A}^{0}-K_\mys{D}(0))/\alpha$ is the spontaneous delay time in the absence of an external force. Eq.~\ref{eqn:model} predicts that the delay time $\taud$ remains close to  $\taud(0)$ for weak forces ($f \ll f_{\mathrm{th}}$). Increasing $f$ results in an increase in $\taud$, which becomes particularly marked when the applied force $f$ approaches the characteristic force $f_{\mathrm{th}}$ of thermal bond rupture. For still larger forces, the delay time saturates at a plateau of $K_\mys{A}^{0}/\alpha$, the largest delay time permitted by our model. In this case where $f \gg f_{\mathrm{th}}$, the lifetime of the gel is extended so much that the rate of strand association approaches zero at $\tw = \taud$, so that any subsequent dissociation event causes an immediate failure.

%%====================================================================
\section{Effect of an external force}

\label{sec:stress}
%%====================================================================

To experimentally check these predictions, we added a small number of monodisperse super-para\-magnetic beads, composed of a cross-linked poly\-styrene with embedded maghemite nanoparticles ($\gamma$-Fe$_{2}$O$_{3}$) (Invitrogen, Dynabeads M-270, radius 1.4 $\mu$m), to a weak depletion gel ($\Delta / \a = 0.62 \pm 0.04$, $\a = 316\pm11$ nm, $\phicolloid = 0.21$, $\cpstar = 2.4$).  Microscopy confirmed that the super-paramagnetic beads were uniformly dispersed throughout the network of the gel. The number density of magnetic beads was estimated as $4.8 \times 10^{14}$ m$^{-3}$, significantly less than the number density $1.5 \times 10^{18}$ m$^{-3}$ of droplets in the gelling system, so the super-paramagnetic beads constitute a trace component. The gel was mounted in a magnetic field gradient $\partial B/\partial z$ generated by a pair of strong permanent NdFeB magnets mounted in a purpose-built alumininium housing \cite{Lin2012}. The strength of the field gradient and consequently the force applied to each super-paramagnetic bead was adjusted by controlling the vertical spacing between the gel and the magnet assembly. The mean spacing $l$ between the magnetic beads in the network was estimated as 17 $\mu$m, which is larger than the correlation length $\R$ of the gel so the forces applied to the gel network by the dilute dispersion of super-paramagnetic particles are expected to be independent of each other and uncorrelated. The force applied to each magnetic tracer particle was calculated from measurements of the in-situ magnetic field $B$, using a Hall-probe magnetometer, and literature estimates \cite{xu_simultaneous_2012} of the intrinsic magnetization properties of the magnetic beads.

\begin{figure}[tbp]
    \begin{center}
    \includegraphics[width=0.9\textwidth]{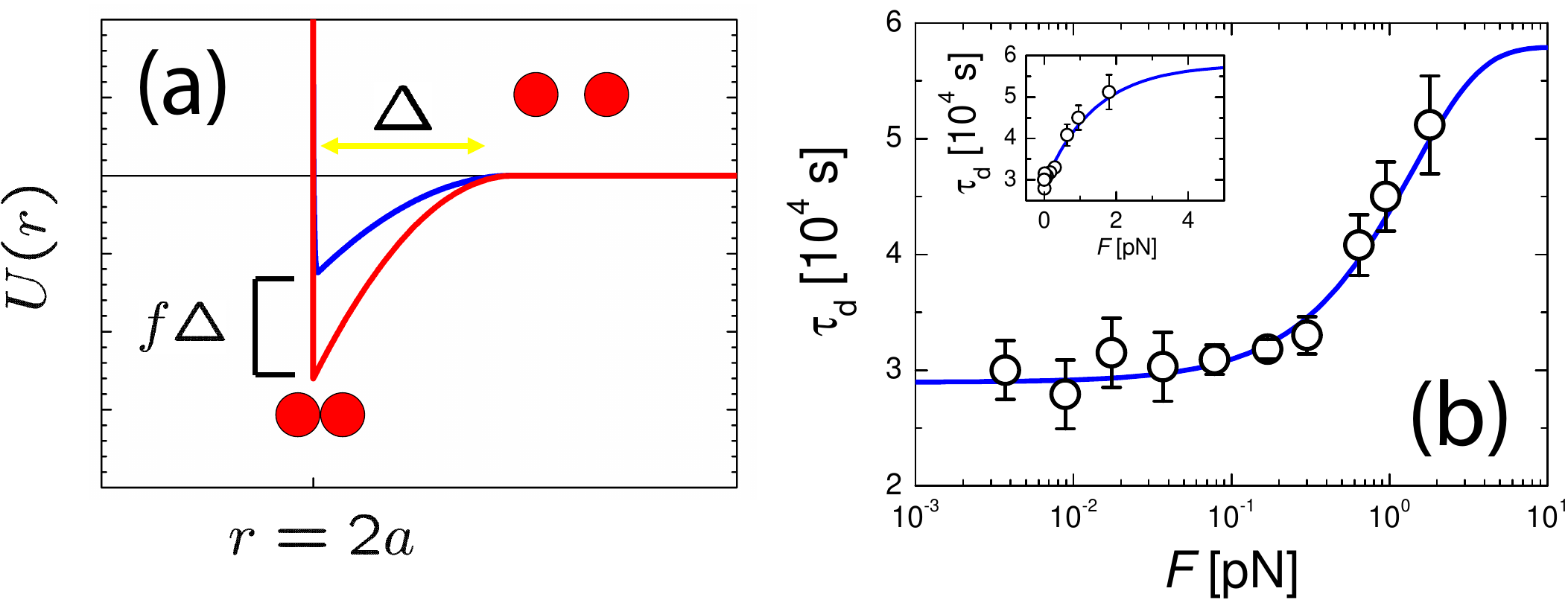}
    \caption{(color online). (a) Shift in the activation barrier for bond dissociation when a force $f$ is applied. $\Delta$ is the range of the attractive interparticle potential. (b) Dependence of the delay time  $\taud$ on the force $F$ generated by super-paramagnetic particles embedded within the network of a gel ($\Delta /\a = 0.62$, $\a =316$ nm,  $\phicolloid = 0.21$, $\cpstar = 2.4$). The force $F$ counteracts the buoyancy force acting on the gel so the delay time  $\taud$ of the gel is increased. The solid line is the dependence predicted by the microscopic model of collapse (Eq.~\ref{eqn:model-fit}), calculated for $\taud(0) = 2.9 \times 10^{4}$ s, $F_{0} = 1.4$ pN, and $\delta = 1$. The inset shows the same data plotted on a linear scale. }%
     \label{fig:potential}
     \end{center}
\end{figure}

Figure~\ref{fig:potential}(b) shows measurements of the delay time of a gel as a function of the force $F$ applied by randomly dispersed super-paramagnetic beads, fixed within the gel network. The delay time increases with $F$ because the external magnetic field $B$ was orientated so that the force generated on the gel by the magnetic particles opposes the buoyancy forces on the gel.  The dependence of $\taud$ on $F$  apparent in Figure~\ref{fig:potential}(b) is in qualitative agreement with the predictions of the microscopic model outlined above, with a significant increase in the delay time being observed for applied forces of $F \approx 1$ pN. At $F \gtrsim 5$ pN however the magnetic particles were observed to be stripped from the gel, as the local yield stress was exceeded and the gel broke. As a result, the high field plateau predicted for the delay time $\taud$ as a function of $F$ (see Section~\ref{sec:model}) could not be detected.

To check quantitatively the microscopic model of Section~\ref{sec:model}, we first note that the force $f$ between particles will be proportional to the applied magnetic force $F$. The constant of proportionality will however depend on the characteristic way in which force propagates through the heterogeneous particle network of a  gel. By analogy with granular materials \cite{majmudar2005contact}, we expect the transmission of force to be highly non-linear.  Forces will propagate predominantly along chains of neighbouring particles while large areas of the surrounding gel remain nearly force-free. If we assume that the contact forces produced by a point force $F$ are concentrated along a linear path of $N_{\mys{c}}$ particles then, in the absence of end effects, the average force on each pair of particles should be $F/N_{\mys{c}}$ where $N_{\mys{c}}$ is the length of the linear force chain. In this case, Eq.~\ref{eqn:model} may be re-expressed as
\begin{equation}\label{eqn:model-fit}
\frac{\taud(F)}{ \taud(0)} = 1 + \delta \left [ 1 - \exp \left (  - \frac{F} {F_{0}}  \right ) \right ]
\end{equation}
where $F_{0} = N f_{\mathrm{th}}$ and $\delta = K_\mys{D}(0)/(K_\mys{A}^{0} - K_\mys{D}(0))$. From the data presented in Fig.~\ref{fig:aging-events} we estimate  $\delta \approx 1$ for the PDMS gel. A non-linear least-squares fit of Eq.~\ref{eqn:model-fit}, with $\delta = 1$,  to the experimental data  (shown as the solid line in Figure~\ref{fig:potential}(b)) yields a good representation of the experimental force distribution. The best-fit parameters were determined as $\taud(0) = 2.90 \pm 0.05 \times 10^{4}$ s and $F_{0} = 1.4 \pm 0.2$ pN. To interpret these values, we note that for $\Delta = 194$ nm the characteristic thermal rupture force is $f_{\mathrm{th}} \approx 0.02$ pN so the number of bonds in the equivalent linear force chain is $N_{\mys{c}} = F_{0}/f_{\mathrm{th}}$ or  $N_{\mys{c}} \approx 70 \pm 10$. $N_{\mys{c}}$ should be comparable to the number of particle bonds between magnetic beads which is $l /(2\a)$ or $\approx$ 30, in the current experiments. This value is indeed comparable to the force chain length found experimentally of $N_{\mys{c}} \approx 70 \pm 10$,  which is very reassuring.  Overall, the force measurements provide reasonably strong support for the microscopic model of gel collapse outlined above. While more work clearly needs to be done, these initial observations strengthen our argument that a catastrophic loss of connectivity as a consequence of thermally-driven dissociation events is ultimately responsible  for delayed collapse.

%%====================================================================
\section{Conclusions}
\label{sec:conclusions}
%%====================================================================

It is useful to briefly summarize what features of the collapse instability in depletion gels are well understood and what is still left to be clarified. Delayed collapse, whereby a gel which is initially sedimentating  slowly after a certain time interval (the \textit{delay} time) switches to a regime of rapid collapse, is well established in systems with long-range attractive potentials. In this paper we have focused predominately on such long-range systems. The distinguishing feature of these gels is that when quenched into a two-phase region the particle network coarsens continuously with time, as a result of thermal breaking and reforming of interparticle bonds. In contrast, the sedimentation kinetics of depletion gels with short-range attractive potentials is more complex, as a result of the interplay between phase separation and dynamical arrest \cite{7444}, and is not as well understood. Finally, we note that the initial particle density is likely to have a significant effect on the time evolution of a gel but the effect of $\phi$ has to date been little studied. 

The delay time $\taud$, which can vary from minutes to months, has been reported to depend on the depth and range of the attractive potential, the volume fraction $\phicolloid$, the particle radius $\a$, the density mismatch  \cite{5926,5389}, the presence of a weak  shear stress  \cite{5382}, and even the width and height of the sample cell \cite{5744,2308}. The link between these effects and the microstructure of a gel is  in many cases currently missing. However the pivotal role of the thermal lifetime of a single particle-particle bond, the Kramers' escape time $\tauesc$, is generally recognized \cite{5237,9091}. Experiments confirm, for instance, that the time required for gel collapse scales approximately linearly with the escape time, over some two orders of magnitude of variation in $\taud$. The large value of the ratio $\taud / \tauesc$, which for the data presented here is of order $\sim 10^{2}$, highlights the significant role the hierarchical structure of a gel plays in the connection between a local bond dissociation event and the eventual overall failure of the network. Time-resolved confocal imaging experiments indicate that, although the gel is continuously coarsening, the most important feature of the thermal restructuring is probably a decrease in connectivity of the network with time. Analysis of the association and dissociation dynamics of strands within the gel suggest that collapse is triggered when the rate of strand association, which is a decreasing function of time, decreases to such an extent that it becomes comparable to the fixed rate of strand dissociation. At this point, the ability of the gel network to support a gravitational stress decreases rapidly and the gel soon afterwards fails macroscopically. This microscopic picture of collapse is supported by preliminary measurements on the effect of an applied point force on the dynamics of collapse. More work on the subtle link between delayed collapse, microscopic dissociation dynamics, and applied force is however essential to fully understand the anomalous mechanical response of weak gels. Particularly puzzling is the role of gravitational stress. The absence of any variation of the delay time with the height of a gel seems, at first sight, to be at variance with the results of the magnetic experiments where $\taud$ increases with applied force. Work is under way to address this intriguing question.

%
%
%%====================================================================
\section{Acknowledgments}
%%====================================================================
%
We gratefully acknowledge support from Bayer CropScience and the UK Engineering and Physical Sciences Research Council. In addition, we thank an anonymous referee for a number of incisive comments which considerably improved an earlier draft.
%
%%====================================================================
%\end{acknowledgments}
%%====================================================================

%\bibliography{dynamics-collapse}
%
%\newpage
%\printnomenclature % turn on nomenclature list

\end{document}